\documentclass[fleqn,usenatbib]{mnras}

\usepackage{newtxtext,newtxmath}
\usepackage[T1]{fontenc}
\usepackage{ae,aecompl}

\usepackage{graphicx}	% Including figure files
\usepackage{amsmath}	% Advanced maths commands
\usepackage{amssymb}	% Extra maths symbols
\usepackage{color}
\usepackage[capitalise]{cleveref}
\usepackage{changepage}
\usepackage{verbatim}
\usepackage{xspace}

\newcommand{\ecms}{erg s$^{-1}$ cm$^{-2}$\xspace}
\newcommand{\molsec}{mol s$^{-1}$\xspace}

\defcitealias{AHearn:1995}{AH95}
\defcitealias{Cochran:2012}{C12}

\title[WHT spectroscopy of comets]{Near-UV and optical spectroscopy of comets using the ISIS spectrograph on the WHT}

\author[]{
M.G. Hyland,$^{1}$\thanks{E-mail: mhyland02@qub.ac.uk}
A. Fitzsimmons,$^{1}$
C. Snodgrass$^{2,3}$
\\
$^{1}$Astrophysics Research Centre, School of Mathematics and Physics, Queen$'$s University Belfast, BT7 1NN, UK\\
$^{2}$Institute for Astronomy, University of Edinburgh, Royal Observatory, Edinburgh EH9 3HJ, UK\\
$^{3}$School of Physical Sciences, The Open University, Walton Hall, Milton Keynes MK7 6AA, UK
}

\date{Accepted XXX. Received YYY; in original form ZZZ}

\pubyear{2019}

\begin{document}
\label{firstpage}
\pagerange{\pageref{firstpage}--\pageref{lastpage}}
\maketitle

% Abstract of the paper
\begin{abstract}
We present an analysis of long-slit cometary spectroscopy using the dual-arm ISIS spectrograph on the 4.2 m WHT. Eleven comets were observed over two nights in 2016 March and we detected the OH (0-0) emission band at 3085 {\AA} in the spectra of five of these comets. Emission bands of the species NH, CN, C$_{3}$, C$_{2}$, NH$_{2}$ and [OI] were also detected. We used Haser modelling to determine molecular production rates and abundance ratios for the observed species. We found that our average abundances relative to OH and CN were generally consistent with those measured in previous studies. 
\end{abstract}

% Select between one and six entries from the list of approved keywords.
% Don't make up new ones.

%%%%%%%%%%%%%%%%%%%%%%%%%%%%%%%%%%%%%%%%%%%%%%%%%%

%%%%%%%%%%%%%%%%% BODY OF PAPER %%%%%%%%%%%%%%%%%%

\section{Introduction}
Comets are minor bodies that are the remnants of planetary formation; therefore by studying the compositions of these bodies we can begin to understand more about the conditions present in the early solar system. Comets become increasingly active with decreasing heliocentric distance, with their activity being driven by the sublimation of ice from the nucleus due to solar radiation. The gases released through sublimation form a transient atmosphere known as the coma and the dust particles released due to the gas pressure gradient form a dust tail \citep{Cochran:2015}. Photometric and spectroscopic observations of comets as they enter the inner solar system allow the identification of the molecular species present in the developing coma.

Previous studies of comets indicate that they have generally similar optical spectra, but the emission band strengths and relative abundances of molecules vary between comets. The CN emission band at 3883{\AA} is typically one of the strongest observed features in the optical spectra of comets \citep{Feldman:2004}. Emission bands due to other species, such as C$_{2}$ and C$_{3}$, are also usually present and measuring the production rates of these species allows abundance ratios to be determined for individual comets. Several studies have already been carried out to investigate relative molecular abundances. \citet[hereafter AH95]{AHearn:1995} have published abundance ratios for 85 comets observed between 1976-1992, while \citet*{Langland:2011} measured abundances in 26 comets observed between 1996-2007. \citet*[hereafter C12]{Cochran:2012} have reported measurements of abundance ratios from observations of 130 comets made at McDonald observatory over a period of thirty years. These surveys find that the majority of observed comets can be broadly classified as one of two types: those that have a `typical' cometary composition and those that are depleted in the carbon-chain molecules C$_{2}$ and C$_{3}$. \citetalias{AHearn:1995} define depleted comets as those having log($\frac{C_{2}}{CN}$) values less than -0.18, while \citetalias{Cochran:2012} define the conditions for depletion as log($\frac{C_{2}}{CN}$) $\leq$ 0.02 and log($\frac{C_{3}}{CN}$) $\leq$ -0.86.

At heliocentric distances less than ${\sim}$3 au activity in comets is dominated by water ice sublimation. Emission from the H$_{2}$O molecules can only be directly observed from the ground in bright comets as the bands are in the near-infrared and sub-millimetre regions, and there is strong absorption in these spectral regions due to the earth's atmosphere \citep{Bockelee-Morvan:2004}. It is therefore often necessary to use the products of the dissociation of H$_{2}$O to investigate the water sublimation rates of comets. Outflowing H$_{2}$O undergoes photodissociation by UV solar photons, with 90 per cent of the water molecules dissociating through the reaction H$_{2}$O + $\gamma$ $\rightarrow$ H + OH \citep{Crovisier:1989}. Water sublimation rates can be measured from ground-based observations using the $^{2}{\Pi}_{3/2}$ J=3/2 ${\Lambda}$-doublet transitions of OH at 18 cm \citep{Crovisier:2002}. OH molecules also have an emission band in the near-UV at 3085 {\AA} due to resonance fluorescence. Measurements of this emission have been made  using the \textit{International Ultraviolet Explorer (IUE)} satellite \citep{Feldman:2004} and from the ground using spectroscopy and narrowband photometric filters \citepalias{AHearn:1995}.

In this paper we describe the analysis of observations of comets made over two nights during 2016 March using the Intermediate dispersion Spectrograph and Imaging System (ISIS) on the 4.2 m William Herschel Telescope (WHT). The blue arm of ISIS provides a rare opportunity to observe the OH molecular emission band at 3085 {\AA} in cometary spectra as many ground-based spectrographs are not sensitive to wavelengths below 3200 {\AA}. Observations of comets performed using ISIS will also contain emission  due to NH, CN, C$_{3}$, C$_{2}$, NH$_{2}$ and [OI].

\section{Observations and reduction}
Observations of several comets were performed on the nights of 2016 March 18 and 19 using the ISIS spectrograph on the WHT on the island of La Palma, Spain. Table~\ref{tab:observation_table} gives the observational log. The D5300 dichroic was used to allow simultaneous observations to be made in the blue and red arms of the spectrograph. The R158B grating was used in the blue arm with a dispersion of 1.62 {\AA} pixel$^{-1}$, and R158R grating was used in the red arm with a dispersion of 1.81 {\AA} pixel$^{-1}$. Each comet exposure was taken using a long slit of width 2.0 arcsec, with a slit length of 3.2 arcmin in the blue arm and 3.5 arcmin in the red arm. All observations were made with the slit oriented at the parallactic angle.The resolving power $R$ in the blue arm was 288, while in the red arm $R$ was 486. The slit width of 2.0 arcsec projected to 9.6 and 8.5 pixels in the blue and red arms respectively. 

Exposures of spectrophotometric flux standard stars were obtained using a 7.0 arcsec wide slit for subsequent flux calibration of the cometary spectra. Observations of the solar analogue star HD28099 were performed on each night to allow the removal of the dust continuum from each cometary spectrum. The comet observations were interspersed with exposures of CuAr and CuNe arc lamps to allow accurate wavelength calibration.

Data reduction was carried out using the NOAO \textsc{iraf} software package. First the images were bias-subtracted and flat-fielded, and twilight sky flats were used to determine the slit illumination function. The arc-lamp exposures were used to wavelength calibrate the standard star and cometary spectra. Cosmic rays were removed from each exposure using the \textsc{lacosmic} package \citep{VanDokkum:2001}. The extraction of each spectrum from its two-dimensional frame was performed using the {\tt apall} task in \textsc{iraf}, which involves summing the flux in an extraction aperture centred on the peak of the spatial profile and identifying regions either side of the peak to use for sky background fitting. Each individual spectrum was inspected and the extraction apertures were chosen to maximise the extracted cometary flux. The sky background regions were selected near the slit edges to ensure that they did not include the outer coma. For comet C/2014 S2 (PANSTARRS) the emission features were found to extend along the length of the slit. The background was therefore removed using an exposure of a region of blank sky $\sim$1 degree from the comet. The one-dimensional cometary spectra were corrected using the atmospheric extinction curve for La Palma \citep{King:1985} and then were flux calibrated using the standard star spectra.

The red-arm spectra required the removal of telluric atmospheric absorption bands. The ESO \textsc{molecfit} package \citep{Smette:2015,Kausch:2015} was used to determine the atmospheric transmission over the relevant wavelength range at each observation time. These transmission functions were then applied to the one-dimensional spectra to remove the telluric features.  

%Log of observations performed using WHT+ISIS on 18 and 19 March 2016.
\begin{table*}
	\centering
	\caption{Observational log.}
	\label{tab:observation_table}
	\begin{tabular}{lcccccccc} 
		\hline
		Comet & UT date$^a$ & r$_{h}^b$ & $\Delta^c$ & $\alpha^d$ & Exp. time$^e$ & Airmass$^f$ & Slit PA$^g$ & Sun-Comet vector$^h$ \\
        & & (au) & (au) & ($\deg$) & (s) & & ($\deg$) & ($\deg$) \\
		\hline
		9P/Tempel 1 & 19.01 & 2.01 & 1.04 & 9.7 & 4x900 & 1.06 & 300.1 & 193.5 \\
        67P/Churyumov-Gerasimenko & 18.97 & 2.61 & 1.63 & 4.9 & 3x900 & 1.13 & 307.0 & 159.9 \\
        77P/Longmore & 19.11 & 2.37 & 1.43 & 10.6 & 4x900 & 1.26 & 355.3 & 290.2 \\
       	81P/Wild 2 & 18.91 & 1.99 & 1.80 & 30.0 & 3x900 & 1.33 & 70.9 & 87.4 \\
     	116P/Wild 4 & 19.17 & 2.24 & 1.60 & 23.2 & 5x900 & 1.67 & 339.6 & 282.7 \\
     	333P/LINEAR & 19.86 & 1.13 & 1.24 & 49.5 & 2x900 & 1.45 & 74.3 & 73.0 \\
     	C/2013 US10 (Catalina) & 18.89 & 2.21 & 2.23 & 25.9 & 2x600 & 1.46 & 98.8 & 69.4 \\
		C/2014 S2 (PANSTARRS) & 19.08 & 2.40 & 1.87 & 22.9 & 2x600 & 1.32 & 189.6 & 197.0 \\
       	C/2014 W2 (PANSTARRS) & 20.18 & 2.67 & 2.73 & 21.2 & 3x900 & 2.03 & 276.9 & 306.0 \\
        C/2015 V2 (Johnson) & 19.96 & 5.30 & 5.12 & 10.8 & 5x900 & 1.40 & 118.8 & 99.6 \\
        P/2016 BA14 (PANSTARRS) & 19.91 & 1.01 & 0.03 & 62.4 & 5x120,8x240 & 1.25 & 15.5 & 86.6 \\
		\hline
	\end{tabular}
    \\
   \footnotesize{$^a$ UT date at start of exposure sequence - 2016 March, $^b$Heliocentric distance, $^c$Geocentric distance, $^d$Phase angle, $^e$Number and length of exposures taken in each sequence, $^f$Airmass at the beginning of the exposure sequence, $^g$Position angle of the slit at the beginning of the exposure sequence, $^h$Position angle of the Sun-Comet vector}
\end{table*}

\section{Results and analysis}
\subsection{Continuum subtraction and measurement of emission fluxes}
Each extracted cometary spectrum was composed of a molecular gas emission spectrum and a reddened solar spectrum due to reflection of sunlight by dust particles. In order to separate the two components the spectrum of the solar analogue star HD28099 was fit to the continuum in each comet spectrum using polynomial fitting and then subtracted, leaving only the gas emission. For the blue-arm spectra, the total flux in each observed emission band was measured using the \textsc{elf} fitting routine in the \textsc{procspec} package in \textsc{idl}\footnote{https://star.pst.qub.ac.uk/${\sim}$rsir/procspec.html}. Table~\ref{tab:blue_flux_table} gives the measured blue-arm fluxes along with the size of the extraction aperture used for each comet.

%Fluxes measured in each emission band using the extracted blue arm spectra.
\begin{table*}
	\centering
	\caption{Integrated fluxes measured using the extracted one-dimensional spectra from the blue arm observations. All data were obtained using a 2 arcsec wide slit.}
	\label{tab:blue_flux_table}
    \begin{adjustwidth}{-1cm}{}
	\begin{tabular}{lcccccc} 
		\hline
		Comet & Slit Length & OH (0,0) flux & NH (0,0) flux & CN (0-0) flux & C$_{3}$ ($\lambda$4050) flux & C$_{2}$ ($\Delta$v=0) flux\\
        & (arcsec) & (\ecms) & (\ecms) & (\ecms) & (\ecms) & (\ecms) \\
		\hline
		9P/Tempel 1 & 17.1 & (1.66${\pm}$0.32)$\times10^{-14}$ &  & (1.04${\pm}$0.04)$\times10^{-14}$ & (6.88${\pm}$0.42)$\times10^{-15}$ & (7.03${\pm}$0.51)$\times10^{-15}$ \\
        67P/Churyumov-Gerasimenko & 5.7 &  &  & (1.04${\pm}$0.08)$\times10^{-15}$ &  &  \\
        77P/Longmore & 17.1 &  &  & (2.65${\pm}$0.19)$\times10^{-15}$ &  & (4.80${\pm}$0.51)$\times10^{-15}$ \\
       	81P/Wild 2 & 3.8 & (2.45${\pm}$0.48)$\times10^{-14}$ &  & (1.87${\pm}$0.04)$\times10^{-14}$ & (7.74${\pm}$0.55)$\times10^{-15}$ & (5.56${\pm}$0.36)$\times10^{-15}$ \\
     	116P/Wild 4 & 17.1 &  &  & (5.19${\pm}$0.18)$\times10^{-15}$ & (2.57${\pm}$0.31)$\times10^{-15}$ & (4.51${\pm}$0.49)$\times10^{-15}$ \\
     	333P/LINEAR & 11.4 & (6.83${\pm}$1.32)$\times10^{-14}$ & (1.55${\pm}$0.08)$\times10^{-14}$ & (7.46${\pm}$0.20)$\times10^{-14}$ & (4.75${\pm}$0.17)$\times10^{-14}$ & (7.02${\pm}$0.18)$\times10^{-14}$ \\
     	C/2013 US10 (Catalina) & 17.1 & (5.23${\pm}$1.00)$\times10^{-13}$ & (9.86${\pm}$0.90)$\times10^{-15}$ & (1.78${\pm}$0.04)$\times10^{-13}$ &  (8.03${\pm}$0.36)$\times10^{-14}$ & (6.75${\pm}$0.26)$\times10^{-14}$ \\
		C/2014 S2 (PANSTARRS) & 17.1 & (2.90${\pm}$0.56)$\times10^{-13}$ & (3.30${\pm}$0.25)$\times10^{-14}$ & (2.25${\pm}$0.06)$\times10^{-13}$ &  (1.42${\pm}$0.06)$\times10^{-13}$ & (1.24${\pm}$0.06)$\times10^{-13}$ \\
       	C/2014 W2 (PANSTARRS) & 17.1 &  &  & (2.70${\pm}$0.05)$\times10^{-14}$ & (1.22${\pm}$0.08)$\times10^{-14}$ & (1.80${\pm}$0.08)$\times10^{-14}$ \\
		\hline
	\end{tabular}
    \end{adjustwidth}
\end{table*}

In the red-arm spectra the NH$_{2}$ (0,3,0)  band was blended with the O(${^1}$D) lines at 6300 {\AA} and 6364 {\AA} due to the low resolution of the spectrograph.  The method outlined in \citet{Fink:2009} was used to determine fluxes for the NH${_2}$  and O(${^1}$D) 6300 {\AA} emission. Note that \citet{Fink:2009} and other investigators traditionally used the linear notation of (0,8,0) $\Pi$ for this NH$_{2}$ band, whereas as pointed out by \citet{Cochran:2012} the bent notation above should be used {\it e.g.} \cite{Ross:1988}.
\citet{Arpigny:1987} used high resolution spectra of comet 1P/Halley to determine that the sum of the NH${_2}$ emission blended with the O(${^1}$D) 6300 {\AA} line and the sum of the lines in the NH${_2}$ 6335 {\AA} peak is in the ratio 0.9:1.0. The pure flux of O(${^1}$D) was therefore calculated by measuring the flux in the 6300 {\AA} blend and subtracting 0.9 times the flux measured in the NH${_2}$ 6335 {\AA} peak. The total flux in the NH${_2}$ (0,3,0) band was obtained by multiplying the 6335 {\AA} flux by 2.0. Table~\ref{tab:red_flux_table} gives these measured fluxes along with the size of the extraction aperture used for each comet.

%Fluxes measured in each emission band using the extracted red arm spectra.
\begin{table*}
	\centering
	\caption{Integrated fluxes measured using the extracted one-dimensional spectra from the red arm observations.}
	\label{tab:red_flux_table}
	\begin{tabular}{lccc} 
		\hline
		Comet & Slit Length & NH$_{2}$ (0,3,0) flux & OI ($\lambda$6300) flux\\
        & (arcsec) & (\ecms) & (\ecms) \\
		\hline
		9P/Tempel 1 & 19.8 & (3.11${\pm}$0.31)$\times10^{-15}$ & (4.10${\pm}$2.19)$\times10^{-16}$ \\
       	81P/Wild 2 & 4.4 & (2.61${\pm}$0.22)$\times10^{-15}$ & (2.18${\pm}$0.22)$\times10^{-15}$ \\
     	C/2013 US10 (Catalina) & 19.8 & (1.24${\pm}$0.11)$\times10^{-14}$ & (8.32${\pm}$0.94)$\times10^{-15}$ \\
		C/2014 S2 (PANSTARRS) & 19.8 & (2.99${\pm}$0.23)$\times10^{-14}$ & (1.28${\pm}$0.19)$\times10^{-14}$ \\
       	C/2014 W2 (PANSTARRS) & 19.8 & (1.18${\pm}$0.21)$\times10^{-15}$ & (8.00${\pm}$1.46)$\times10^{-16}$ \\
		\hline
	\end{tabular}
\end{table*}

The uncertainty on the measured flux was determined by adding the relative uncertainties of three components in quadrature. The first component was the uncertainty in the flux of the spectrophotometric standard star used (${\sim}$1 per cent, \citet{Bohlin:2007}). The other two components were the instrument calibration uncertainty at the emission band wavelength and the uncertainty from the \textsc{elf} fitting routine respectively. 

\subsection{Observed gas emission bands}
\begin{figure}
	\includegraphics[width=\columnwidth]{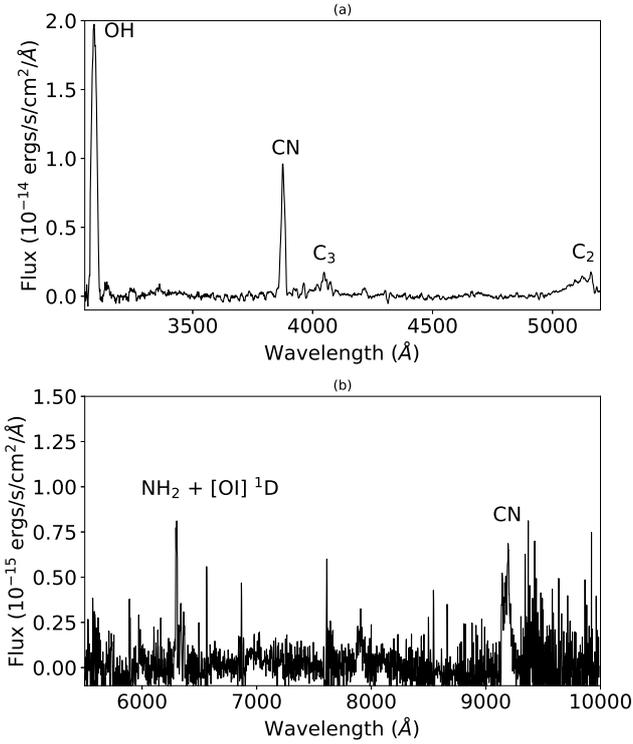}
    \caption{(a) Extracted blue arm dust-subtracted spectrum of comet C/2013 US10 and (b) extracted red arm dust-subtracted spectrum. The molecular gas emission bands are labelled. Note the different y-axis scales in each panel.}
   \label{fig:us10_spec}
\end{figure}

\begin{figure}
	\includegraphics[width=\columnwidth]{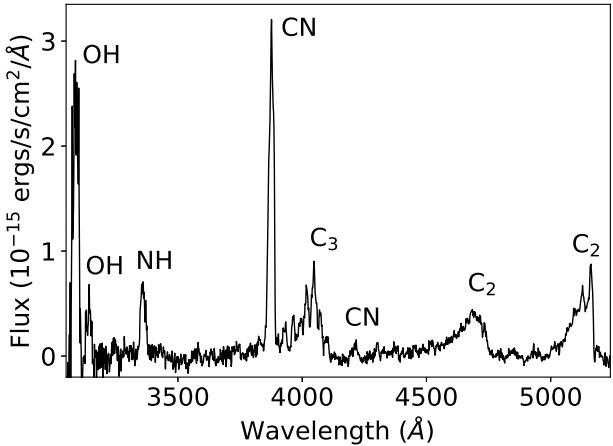}
    \caption{Extracted dust-subtracted blue arm spectrum of comet 333P/LINEAR. The molecular gas emission bands are labelled.}
   \label{fig:333p_blue_spec}
\end{figure}

After removing the dust continuum we assessed the gas emission bands that were present in each spectrum. The near-UV OH emission band at 3085 {\AA} was detected in the blue-arm spectra of five of the comets, as well as emission bands due to NH, CN, C$_{3}$ and C$_{2}$. These comets were 9P/Tempel 1, 81P/Wild 2, 333P/LINEAR, C/2013 US10 (Catalina) and C/2014 S2 (PANSTARRS). The red-arm spectra of these comets contained emission bands due to NH$_{2}$ and CN as well as O($^{1}$D) emission features. \cref{fig:us10_spec,fig:333p_blue_spec} show the dust-subtracted spectra of comets C/2013 US10 and 333P/LINEAR with the observed emission bands labelled.

%C/2013 US10 extracted blue and red arm spectra.

%333P/LINEAR extracted blue arm spectrum.

\begin{figure}
	\includegraphics[width=\columnwidth]{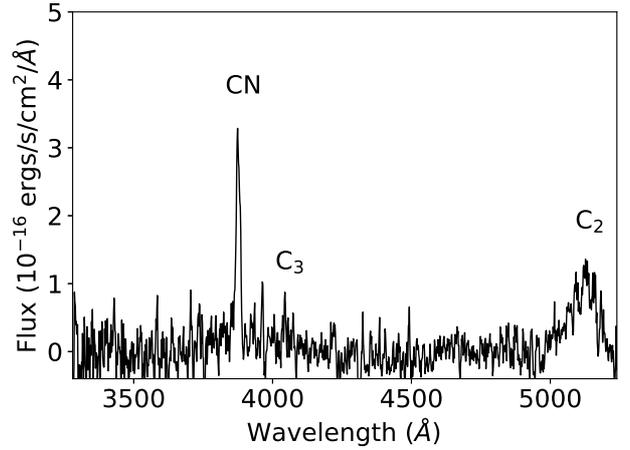}
    \caption{Extracted dust-subtracted blue arm spectrum of  comet 116P/Wild 4. The molecular gas emission bands are labelled.}
   \label{fig:116P_blue_spec}
\end{figure}

The 3085 {\AA} OH band was not detected in the spectra of  comets 116P/Wild 4 and C/2014 W2 (PANSTARRS), but emission bands due to CN, C$_{3}$ and C$_{2}$ were present. The spectrum of comet 77P/Longmore contained emission due to CN and C$_{2}$, while only CN was detected in the spectrum of comet 67P/Churyumov-Gerasimenko. Fig.~\ref{fig:116P_blue_spec} shows the dust-subtracted spectrum of comet 116P/Wild 4 with the observed emission bands labelled.

%116P/Wild 4 extracted blue arm spectrum.

\begin{figure}
	\includegraphics[width=\columnwidth]{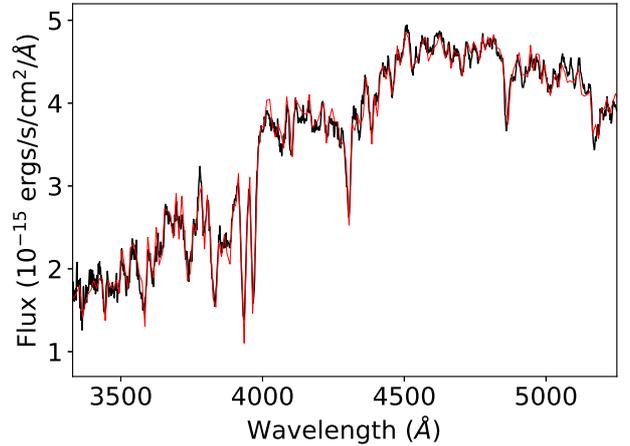}
    \caption{Extracted spectrum of  cometary fragment P/2016 BA14. The fitted solar spectrum is plotted in red.}
   \label{fig:BA14_blue_spec}
\end{figure}

The remaining comets, C/2015 V2 (Johnson) and P/2016 BA14 (PANSTARRS), did not exhibit any detectable gas emission, as shown in Fig.~\ref{fig:BA14_blue_spec}. P/2016 BA14 is thought to share a parent body with  comet 252P/LINEAR. It was of particular interest at the time of our observations as this object made a close approach to Earth several days later (${\Delta}$=0.02 au on 2016 March 22). Unfortunately it was not possible to observe the parent comet 252P/LINEAR from La Palma at the time of our observations.

%P/2016 BA14 extracted blue arm spectrum.

\subsection{Haser modelling}
The two dimensional spectra contained information about the spatial distribution of the gas molecules in the coma of each comet. Spatial profiles for the gas emission features were extracted from the sky background-subtracted comet exposures by defining a range of pixels along the dispersion axis that contained the emission band of a particular species and measuring the total flux in these pixels at each position along the spatial direction. The contribution of the dust to this total flux was removed by extracting a spatial profile from a region that contained no gas emission features, with the flux in that region being only due to reflection by dust in the coma. The fitted one-dimensional dust spectrum was used to determine the scaling factor for the dust profile by measuring the relative strength of the continuum in the centre of the emission band compared to that in the centre of the dust region. The scaled dust profile was subtracted from the emission profile to leave the contribution due to the gas only. 

The measured spatial fluxes were converted to column densities using the fluorescence efficiencies for each species. The fluorescence efficiencies for the OH (0-0), NH (0-0) and CN (0-0) features are strongly dependent on the heliocentric velocity of the comet due to the Swings effect \citep{Swings:1941}, and were determined using D. Schleicher's online tool\footnote{http://asteroid.lowell.edu/comet/cover{\_}gfactor.html}. The C$_{3}$ ($\lambda$4050) and the C$_{2}$ ($\Delta$v=0) fluorescence efficiencies were taken from \citetalias{AHearn:1995} and \citetalias{Cochran:2012}, while the NH$_{2}$ (0,3,0) fluorescence efficiency was that used by \citet{Fink:2009}. The fluorescence efficiencies were scaled as $r_{h}^{-2}$.

The Haser model \citep{Haser:1957} describes the photodissociation of parent molecules into daughter molecules in a spherically symmetric coma. The column density of the daughter molecules $N_{d}$ as a function of distance from the nucleus $\rho$ for a molecular production rate $Q$ is given by the equation
\begin{equation}
N_{d}(\rho) = \frac{Q}{2{\pi}{\rho}v}\frac{{l}_{d}}{{l}_{p} - {l}_{d}}\left(\int_{0}^{{\rho}/{l}_{p}}K_{0}(y)\,dy - \int_{0}^{{\rho}/{l}_{d}}K_{0}(y)\,dy \right) \label{eq:1}
\end{equation}
where $K_{0}(y)$ is a modified Bessel function of the second kind. The quantities $l_{p}$ and $l_{d}$ are the scale lengths of the parent and daughter molecules respectively, and $v$ is the radial outflow velocity which is calculated using the equation from \citet{Cochran:1993}:
\begin{equation}
v \, (km \, s^{-1}) = 0.85r_{h}^{-0.5} \label{eq:2}
\end{equation}

A model Haser profile was fit to the measured column density profile of each observed species to determine its production rate. The nucleus was placed at the centre of the slit for each exposure, and therefore the extracted profiles have components at $\rho$ values either side of the nucleus. As the Haser model assumes spherical symmetry in the coma, the profiles in the two directions were averaged and the production rates were determined using these average profiles. A series of model profiles for different values of the production rate were generated, and the model profile where the $\chi^{2}$ statistic was minimised gave the best-fitting production rate. The additional uncertainty in $Q$ was determined by finding the difference between the best-fitting production rate and the model production rate where $\chi^{2} = \chi^{2}_{min} + 1$ \citep{Hughes:2010}.

We measured the molecular production rates using scale lengths from two sources, which are given in Table~\ref{tab:scalelength_table}. The production rates in Table~\ref{tab:ahearn_blue_prod_rate_table} were determined using the scale lengths from \citetalias{AHearn:1995}, while those in Table~\ref{tab:cochran_prod_rate_table} used the scale lengths from \citetalias{Cochran:2012}. We chose to use the same set of fixed scale lengths as previous authors to allow direct comparison of our abundance ratio measurements with the ratios determined in those studies. The parent and daughter scale lengths are usually scaled as $r_{h}^{2}$, but \citet{Langland:2011} note that the theoretical scale length variation should instead be $r_{h}^{1.5}$. Power law fitting can be carried out to measure the heliocentric distance dependence for the scale lengths of each species from the observations themselves, however in this work we did not have sufficient telescope time to observe each comet over a range of $r_{h}$ values. For consistency with previous studies we used a heliocentric distance scaling of $r_{h}^{2}$, the exception being the parent  of C$_{2}$ which has been found to scale as $r_{h}^{2.5}$ \citep{Cochran:1985}.

%Scale lengths used in Haser model fitting - from A'Hearn et al. (1995), Cochran et al. (2012).
\begin{table*}
	\centering
	\caption{Scale lengths used in the Haser model fitting to determine production rates for the observed species.}
	\label{tab:scalelength_table}
	\begin{tabular}{lcccc} 
		\hline
          & \multicolumn{2}{c}{\citetalias{AHearn:1995} scale lengths} & \multicolumn{2}{c}{\citetalias{Cochran:2012} scale lengths} \\
        Emission feature & l$_{p}$ $^a$ (km) & l$_{d}$ $^a$ (km) & l$_{p}$ $^b$ (km) & l$_{d}$ $^b$ (km) \\
		\hline
		OH (0-0) & 2.4$\times10^{4}$ & 1.6$\times10^{5}$ & 2.4$\times10^{4}$ & 1.6$\times10^{5}$ \\
       	NH (0-0) & 5.0$\times10^{4}$ & 1.5$\times10^{5}$ & 5.0$\times10^{4}$ & 1.5$\times10^{5}$ \\
     	CN (0-0) & 1.3$\times10^{4}$ & 2.1$\times10^{5}$ & 1.7$\times10^{4}$ & 3.0$\times10^{5}$ \\
     	C$_{3}$ ($\lambda$4050) & 2.8$\times10^{3}$ & 2.7$\times10^{4}$ & 3.1$\times10^{3}$ & 1.5$\times10^{5}$ \\
     	C$_{2}$ ($\Delta$v=0) & 2.2$\times10^{4}$ & 6.6$\times10^{4}$ & 2.5$\times10^{4}$ & 1.2$\times10^{5}$ \\
		NH$_{2}$ (0,3,0) & - & - & 4.1$\times10^{3}$ & 6.2$\times10^{4}$ \\
		\hline
	\end{tabular}
    \\
    \footnotesize{$^a$Scales as r$_{h}^{2}$, $^b$Scales as r$_{h}^{2}$, except the C$_{2}$ parent which scales as r$_{h}^{2.5}$}
\end{table*}

%Production rates determined using the A'Hearn et al. (1995) scale lengths.
\begin{table*}
	\centering
	\caption{Production rates determined using the \citetalias{AHearn:1995} scale lengths.}
	\label{tab:ahearn_blue_prod_rate_table}
    \begin{adjustwidth}{-1cm}{}
	\begin{tabular}{lcccccc} 
		\hline
		Comet & OH (0-0) & NH (0-0) & CN (0-0) & C$_{3}$ ($\lambda$4050) & C$_{2}$ ($\Delta$v=0) & H$_{2}$O$^a$ \\
        & (\molsec) & (\molsec) & (\molsec) & (\molsec) & (\molsec) & (\molsec) \\
		\hline
		9P/Tempel 1 & (3.25${\pm}$0.63)$\times10^{26}$ &  & (6.97${\pm}$0.21)$\times10^{23}$ & (1.28${\pm}$0.04)$\times10^{23}$ & (8.07${\pm}$0.58)$\times10^{23}$ & (3.61${\pm}$0.70)$\times10^{26}$ \\
        67P/Churyumov-Gerasimenko &  &  & (5.12${\pm}$0.40)$\times10^{23}$ &  &  &  \\
        77P/Longmore &  &  & (5.05${\pm}$0.36)$\times10^{23}$ &  & (9.79${\pm}$1.03)$\times10^{23}$ &  \\
       	81P/Wild 2 & (1.31${\pm}$0.25)$\times10^{27}$ &  & (4.79${\pm}$0.08)$\times10^{24}$ & (2.33${\pm}$0.10)$\times10^{23}$ & (2.39${\pm}$0.16)$\times10^{24}$ & (1.46${\pm}$0.28)$\times10^{27}$ \\
     	116P/Wild 4 &  &  & (4.65${\pm}$0.10)$\times10^{23}$ & (6.22${\pm}$0.54)$\times10^{22}$ & (7.98${\pm}$0.86)$\times10^{23}$ &  \\
     	333P/LINEAR & (3.88${\pm}$0.77)$\times10^{26}$ & (3.01${\pm}$0.17)$\times10^{24}$ & (1.77${\pm}$0.03)$\times10^{24}$ & (1.26${\pm}$0.04)$\times10^{23}$ & (2.10${\pm}$0.05)$\times10^{24}$ & (4.31${\pm}$0.86)$\times10^{26}$ \\
     	C/2013 US10 (Catalina) & (7.94${\pm}$1.50)$\times10^{27}$ & (1.22${\pm}$0.09)$\times10^{25}$ & (1.77${\pm}$0.03)$\times10^{25}$ & (1.18${\pm}$0.03)$\times10^{24}$ & (1.25${\pm}$0.05)$\times10^{25}$ & (8.82${\pm}$1.67)$\times10^{27}$ \\
		C/2014 S2 (PANSTARRS) & (8.50${\pm}$1.62)$\times10^{27}$ & (5.22${\pm}$0.22)$\times10^{25}$ & (2.88${\pm}$0.04)$\times10^{25}$ & (2.76${\pm}$0.07)$\times10^{24}$ & (2.80${\pm}$0.13)$\times10^{25}$ & (9.44${\pm}$1.80)$\times10^{27}$ \\
       	C/2014 W2 (PANSTARRS) &  &  & (5.25${\pm}$0.10)$\times10^{24}$ & (6.95${\pm}$0.20)$\times10^{23}$ & (6.17${\pm}$0.29)$\times10^{24}$ &  \\
		\hline
	\end{tabular}
    \end{adjustwidth}
    \footnotesize{$^a$Calculated from the OH (0-0) production rate assuming a 90 per cent branching ratio for the photodissociation of H$_{2}$O \citep{Crovisier:1989}.}
\end{table*}

%Production rates determined using the Cochran et al. (2012) scale lengths.
\begin{table*}
	\centering
	\caption{Production rates determined using the \citetalias{Cochran:2012} scale lengths. The OH (0-0) and NH (0-0) parent and daughter scale lengths are the same as those used in \citetalias{AHearn:1995}, and their measured production rates are not repeated here.}
	\label{tab:cochran_prod_rate_table}
	\begin{tabular}{lcccc} 
		\hline
		Comet & CN (0-0) & C$_{3}$ ($\lambda$4050) & C$_{2}$ ($\Delta$v=0) & NH$_{2}$ (0,3,0) \\
        & (\molsec) & (\molsec) & (\molsec) & (\molsec) \\
		\hline
		9P/Tempel 1 & (8.14${\pm}$0.25)$\times10^{23}$ & (3.17${\pm}$0.11)$\times10^{23}$ & (1.16${\pm}$0.08)$\times10^{24}$ & (5.16${\pm}$0.51)$\times10^{24}$ \\
        67P/Churyumov-Gerasimenko & (6.34${\pm}$0.49)$\times10^{23}$ &  &  &  \\
        77P/Longmore & (6.17${\pm}$0.44)$\times10^{23}$ &  & (1.50${\pm}$0.16)$\times10^{24}$ &  \\
       	81P/Wild 2 & (5.51${\pm}$0.09)$\times10^{24}$ & (5.43${\pm}$0.23)$\times10^{23}$ & (3.46${\pm}$0.22)$\times10^{24}$ & (1.50${\pm}$0.13)$\times10^{25}$ \\
     	116P/Wild 4 & (5.44${\pm}$0.13)$\times10^{23}$ & (1.42${\pm}$0.12)$\times10^{23}$ & (1.19${\pm}$0.13)$\times10^{24}$ &  \\
     	333P/LINEAR & (1.84${\pm}$0.03)$\times10^{24}$ & (2.86${\pm}$0.09)$\times10^{23}$ & (2.31${\pm}$0.06)$\times10^{24}$ &  \\
     	C/2013 US10 (Catalina) & (2.00${\pm}$0.03)$\times10^{25}$ & (2.67${\pm}$0.07)$\times10^{24}$ & (1.84${\pm}$0.07)$\times10^{25}$ & (3.60${\pm}$0.31)$\times10^{25}$ \\
		C/2014 S2 (PANSTARRS) & (3.19${\pm}$0.05)$\times10^{25}$ & (6.35${\pm}$0.15)$\times10^{24}$ & (4.30${\pm}$0.20)$\times10^{25}$ & (1.00${\pm}$0.08)$\times10^{26}$ \\
       	C/2014 W2 (PANSTARRS) & (6.12${\pm}$0.11)$\times10^{24}$ & (1.68${\pm}$0.05)$\times10^{24}$ & (9.84${\pm}$0.46)$\times10^{24}$ & (6.20${\pm}$1.12)$\times10^{24}$ \\
		\hline
	\end{tabular}
\end{table*}

\begin{figure*}
	\includegraphics{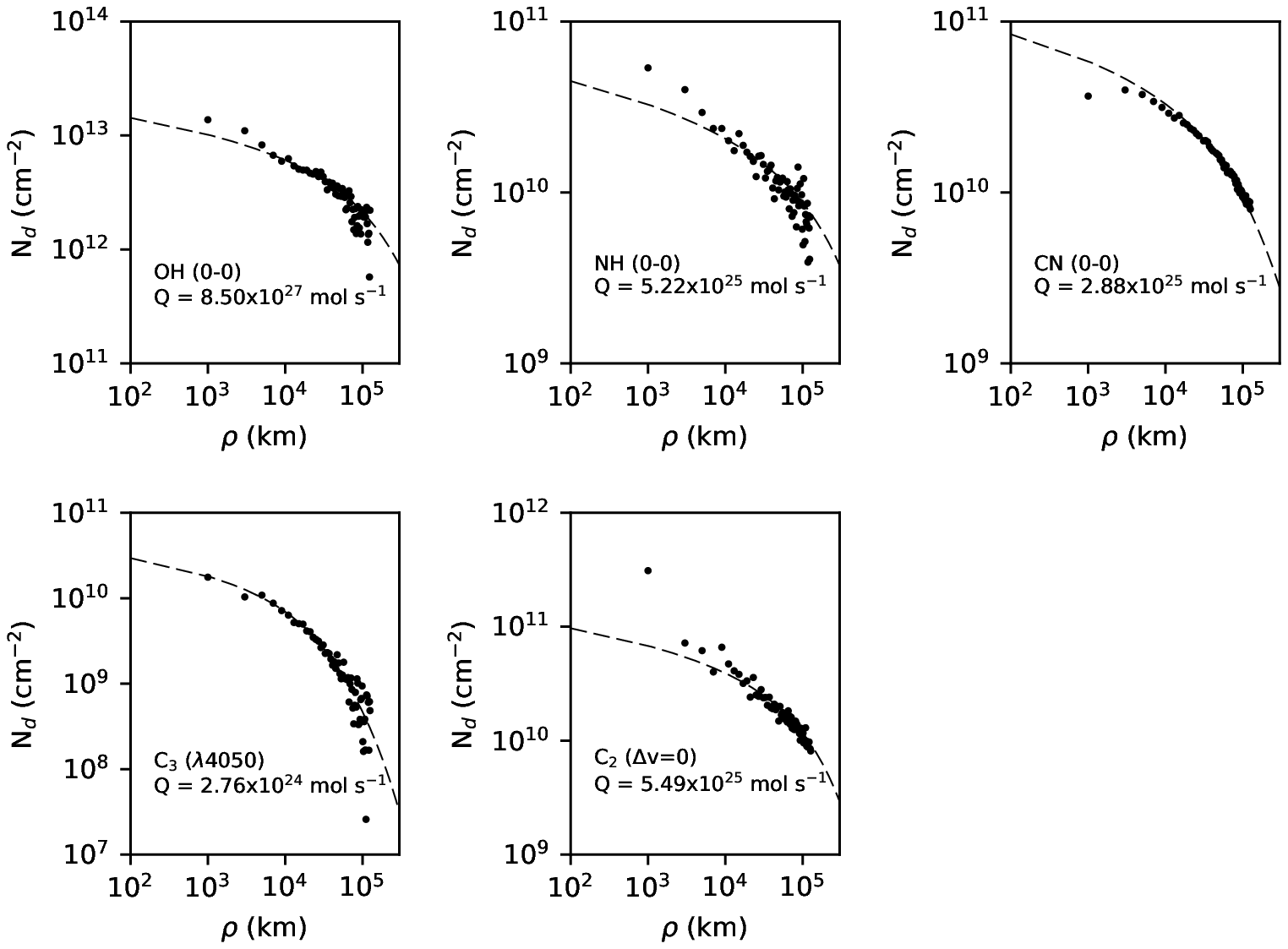}
    \caption{The extracted spatial profiles for  comet C/2014 S2. The gas emission for this comet extends over the length of the slit. The production rates determined using the \citetalias{AHearn:1995} scale lengths are given for each emission feature, and the corresponding Haser profile for that production rate is plotted as a dashed line.}
   \label{fig:s2_profiles_ahearn}
\end{figure*}

\begin{figure*}
	\includegraphics{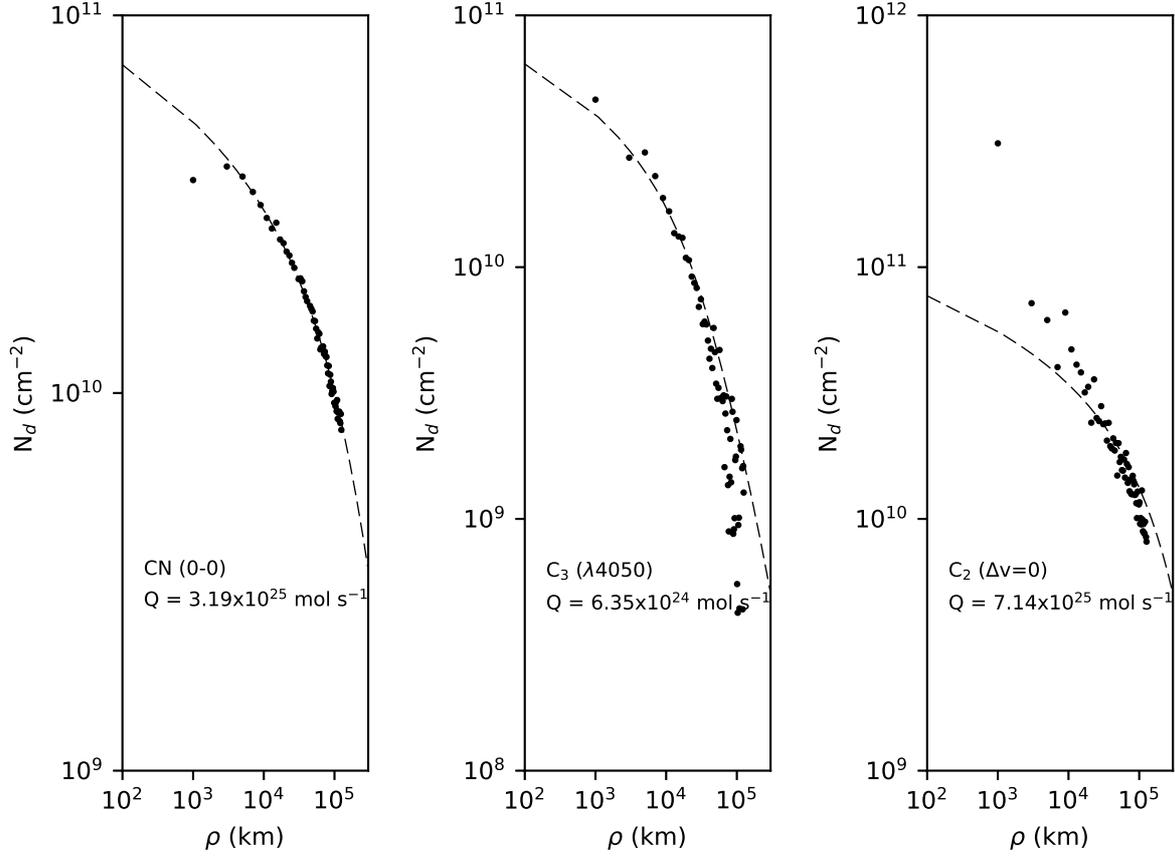}
    \caption{The extracted spatial profiles for comet C/2014 S2. The production rates determined using the \citetalias{Cochran:2012} scale lengths are given for each emission feature, and the corresponding Haser profile for that production rate is plotted as a dashed line. The OH and NH scale lengths are the same as those in \citetalias{AHearn:1995} and the plots for these features are not repeated here.}
   \label{fig:s2_profiles_cochran}
\end{figure*}

Fig.~\ref{fig:s2_profiles_ahearn} shows the spatial profiles that were extracted for comet C/2014 S2. The production rates determined using the \citetalias{AHearn:1995} scale lengths are given for each emission feature, and the corresponding Haser profile for that production rate is plotted as a dashed line. Fig.~\ref{fig:s2_profiles_cochran} shows the fits to the spatial profiles of C/2014 S2 using the scale lengths from \citetalias{Cochran:2012}. Table~\ref{tab:all_ratio_table} and Table~\ref{tab:all_ratio_table2} give the measured abundance ratios for each comet determined using the scale lengths in \citetalias{AHearn:1995} and \citetalias{Cochran:2012}, along with the weighted average and standard deviation for each ratio. Due to the existence of carbon-chain depleted comets, we have included two measurements for the average abundances of C$_{2}$ and C$_{3}$: one value that includes all the measurements and a second value that only includes the comets that fit the definition of `typical'. The average abundance ratios determined by \citetalias{AHearn:1995} and \citetalias{Cochran:2012} for the `typical' and `depleted' comets are also given in the tables for comparison.

%Extracted spatial profiles for C/2014 S2 - A'Hearn scale lengths.

%Extracted spatial profiles for C/2014 S2 - Cochran scale lengths.

%Abundances measured relative to OH.
\begin{table*}
	\centering
	\caption{Measured abundance ratios relative to OH. The labels \citetalias{AHearn:1995} and \citetalias{Cochran:2012} refer to the source of the scale lengths used to derive the production rates. The average abundance ratios for `typical' and `depleted' comets determined in \citetalias{AHearn:1995} and \citetalias{Cochran:2012} are included for comparison.}
	\label{tab:all_ratio_table}
    \begin{adjustwidth}{-1.2cm}{}
	\begin{tabular}{lccccccccc} 
		\hline
		Comet & \multicolumn{2}{c}{log($\frac{\text{CN}}{\text{OH}}$)}  & \multicolumn{2}{c}{log($\frac{\text{C$_{2}$}}{\text{OH}}$)} & \multicolumn{2}{c}{log($\frac{\text{C$_{3}$}}{\text{OH}}$)} & \multicolumn{2}{c}{log($\frac{\text{NH}}{\text{OH}}$)} & log($\frac{\text{NH$_{2}$}}{\text{OH}}$)\\
        & \citetalias{AHearn:1995} & \citetalias{Cochran:2012} & \citetalias{AHearn:1995} & \citetalias{Cochran:2012} & \citetalias{AHearn:1995} & \citetalias{Cochran:2012} & \citetalias{AHearn:1995} & \citetalias{Cochran:2012} & \citetalias{Cochran:2012}\\
		\hline
		9P/Tempel 1 & -2.67${\pm}$0.08 & -2.60${\pm}$0.08 & -2.61${\pm}$0.09 & -2.45${\pm}$0.09 & -3.40${\pm}$0.09 & -3.01${\pm}$0.09 &  &  & -1.80${\pm}$0.09\\
       	81P/Wild 2 & -2.44${\pm}$0.08 & -2.38${\pm}$0.08 & -2.74${\pm}$0.09 & -2.58${\pm}$0.09 & -3.75${\pm}$0.09 & -3.38${\pm}$0.09 &  &  & -1.94${\pm}$0.09\\
     	333P/LINEAR & -2.34${\pm}$0.09 & -2.32${\pm}$0.09 & -2.27${\pm}$0.09 & -2.23${\pm}$0.09 & -3.49${\pm}$0.09 & -3.13${\pm}$0.09 & -2.11${\pm}$0.09 & -2.11${\pm}$0.09 & -\\
     	C/2013 US10 (Catalina) & -2.65${\pm}$0.08 & -2.60${\pm}$0.08 & -2.80${\pm}$0.08 & -2.64${\pm}$0.08 & -3.83${\pm}$0.08 & -3.47${\pm}$0.08 & -2.81${\pm}$0.09 & -2.81${\pm}$0.09 & -2.34${\pm}$0.09\\
		C/2014 S2 (PANSTARRS) & -2.48${\pm}$0.08 & -2.43${\pm}$0.08 & -2.48${\pm}$0.09 & -2.30${\pm}$0.09 & -3.49${\pm}$0.08 & -3.13${\pm}$0.08 & -2.21${\pm}$0.08 & -2.21${\pm}$0.08 & -1.93${\pm}$0.09\\
		\hline
        Measured average & -2.52${\pm}$0.13 & -2.47${\pm}$0.11 & -2.58${\pm}$0.19 & -2.43${\pm}$0.16 & -3.59${\pm}$0.17 & -3.23${\pm}$0.17 & -2.37${\pm}$0.31 & -2.37${\pm}$0.31 & -2.01${\pm}$0.20\\
        Measured average - `typical' & & & -2.45${\pm}$0.14 & -2.32${\pm}$0.09 & -3.46${\pm}$0.04 & -3.09${\pm}$0.06 & & & \\
        \\
        Published average - `typical' & -2.50${\pm}$0.18 & -2.29${\pm}$0.33 & -2.44${\pm}$0.20 & -2.13${\pm}$0.36 & -3.59${\pm}$0.29 & -2.96${\pm}$0.36 & -2.37${\pm}$0.27 & -1.81${\pm}$0.39 & -2.38${\pm}$0.42\\
        Published average - `depleted' & -2.69${\pm}$0.14 & -2.19${\pm}$0.16 & -3.30${\pm}$0.35 & -2.43${\pm}$0.62 & -4.18${\pm}$0.28 & -3.25${\pm}$0.42 & -2.48${\pm}$0.34 & -1.75${\pm}$0.32 & -2.18\\
        \hline
	\end{tabular}
    \end{adjustwidth}
\end{table*}

%Abundances measured relative to CN.
\begin{table*}
	\centering
	\caption{Measured abundance ratios relative to CN. The labels \citetalias{AHearn:1995} and \citetalias{Cochran:2012} refer to the source of the scale lengths used to derive the production rates. The average abundance ratios for `typical' and `depleted' comets determined in \citetalias{AHearn:1995} and \citetalias{Cochran:2012} are included for comparison.}
	\label{tab:all_ratio_table2}
    \begin{adjustwidth}{-1cm}{}
	\begin{tabular}{lccccccccc} 
		\hline
		Comet & \multicolumn{2}{c}{log($\frac{\text{OH}}{\text{CN}}$)}  & \multicolumn{2}{c}{log($\frac{\text{C$_{2}$}}{\text{CN}}$)} & \multicolumn{2}{c}{log($\frac{\text{C$_{3}$}}{\text{CN}}$)} & \multicolumn{2}{c}{log($\frac{\text{NH}}{\text{CN}}$)} & log($\frac{\text{NH$_{2}$}}{\text{CN}}$) \\
        & \citetalias{AHearn:1995} & \citetalias{Cochran:2012} & \citetalias{AHearn:1995} & \citetalias{Cochran:2012} & \citetalias{AHearn:1995} & \citetalias{Cochran:2012} & \citetalias{AHearn:1995} & \citetalias{Cochran:2012} & \citetalias{Cochran:2012} \\
		\hline
		9P/Tempel 1 & 2.67${\pm}$0.08 & 2.60${\pm}$0.08 & 0.06${\pm}$0.03 & 0.15${\pm}$0.03 & -0.74${\pm}$0.02 & -0.41${\pm}$0.02 &  &  & 0.80${\pm}$0.04 \\
        77P/Longmore & & & 0.29${\pm}$0.06 & 0.39${\pm}$0.06 & & & & & \\
       	81P/Wild 2 & 2.44${\pm}$0.08 & 2.38${\pm}$0.08 & -0.30${\pm}$0.03 & -0.20${\pm}$0.03 & -1.31${\pm}$0.02 & -1.01${\pm}$0.02 &  &  & 0.43${\pm}$0.04 \\     	
		116P/Wild 4 &  &  & 0.23${\pm}$0.05 & 0.34${\pm}$0.05 & -0.87${\pm}$0.04 & -0.58${\pm}$0.04 &  &  &  \\
     	333P/LINEAR & 2.34${\pm}$0.09 & 2.32${\pm}$0.09 & 0.07${\pm}$0.01 & 0.10${\pm}$0.01 & -1.15${\pm}$0.02 & -0.81${\pm}$0.02 & 0.23${\pm}$0.03 & 0.24${\pm}$0.03 &  \\
     	C/2013 US10 (Catalina) & 2.65${\pm}$0.08 & 2.60${\pm}$0.08 & -0.15${\pm}$0.02 & -0.04${\pm}$0.02 & -1.18${\pm}$0.01 & -0.87${\pm}$0.01 & -0.16${\pm}$0.03 & -0.21${\pm}$0.03 & 0.26${\pm}$0.04 \\
		C/2014 S2 (PANSTARRS) & 2.48${\pm}$0.08 & 2.43${\pm}$0.08 & -0.01${\pm}$0.02 & 0.13${\pm}$0.02 & -1.01${\pm}$0.01 & -0.70${\pm}$0.01 & 0.26${\pm}$0.02 & 0.21${\pm}$0.02 & 0.50${\pm}$0.03 \\
       	C/2014 W2 (PANSTARRS) &  &  & 0.07${\pm}$0.02 & 0.21${\pm}$0.02 & -0.88${\pm}$0.01 & -0.56${\pm}$0.01 &  &  & 0.01${\pm}$0.08 \\
		\hline
        Measured average & 2.52${\pm}$0.13 & 2.47${\pm}$0.11 & 0.01${\pm}$0.16 & 0.09${\pm}$0.15 & -1.04${\pm}$0.17 & -0.72${\pm}$0.17 & 0.18${\pm}$0.16 & 0.15${\pm}$0.17 & 0.45${\pm}$0.20 \\
        Measured average - `typical' & & & 0.07${\pm}$0.06 & 0.14${\pm}$0.07 & -0.96${\pm}$0.13 & -0.64${\pm}$0.13 & & & \\
        \\
        Published average - `typical' & 2.50${\pm}$0.18 & 2.29${\pm}$0.33 & 0.06${\pm}$0.10 & 0.16${\pm}$0.14 & -1.09${\pm}$0.34 & -0.67${\pm}$0.15 & 0.13${\pm}$0.32 & 0.48${\pm}$0.21 & -0.09${\pm}$0.26 \\
        Published average - `depleted' & 2.69${\pm}$0.14 & 2.19${\pm}$0.16 & -0.61${\pm}$0.35 & -0.24${\pm}$0.60 & -1.49${\pm}$0.31 & -1.06${\pm}$0.39 & 0.21${\pm}$0.37 & 0.44${\pm}$0.28 & 0.01 \\
        \hline       
	\end{tabular}
    \end{adjustwidth}
\end{table*}

The production rates for NH${_2}$ and H${_2}$O (using the O(${^1}$D) 6300 {\AA} emission feature) were calculated using their integrated fluxes due to to the difficulty of extracting spatial profiles from the red-arm exposures. The production rate for NH${_2}$ was determined by comparing the measured emission band flux to that obtained by integrating a Haser model profile over the dimensions of the extraction aperture. The NH$_{2}$ production rates were determined using the scale lengths in \citet{Fink:2009} and \citetalias{Cochran:2012} as \citetalias{AHearn:1995} do not observe NH$_{2}$ emission in their study. 

\subsection{\texorpdfstring{O(${^1}$D) 6300 {\AA} emission}{O1D}}
The O(${^1}$D) 6300 {\AA} emission is not due to resonance fluorescence, but is produced when H${_2}$O molecules undergo photodissociation to directly form oxygen in the $^{1}$D state. This feature has been used in other studies to measure cometary water production rates (e.g. \citet{Fink:2009,McKay:2018}). 

A Haser correction factor was applied to the measured O(${^1}$D) 6300 {\AA} flux to account for the fact that only a small portion of the coma was observed through the slit. The flux was converted to a luminosity $L_{OI}$ and the H${_2}$O production rate $Q_{H{_2}O}$ was calculated using the equation from \citet{Fink:2009}:
\begin{equation}
Q_{H{_2}O} \, (mol \, s^{-1}) = 16.0L_{OI} \, (photons\, s^{-1}) 
\end{equation} 

We found that the H$_{2}$O production rates determined using the O(${^1}$D) 6300 {\AA} feature were on average an order of magnitude higher than those derived from the OH emission. Our measured average for the $\frac{NH_{2}}{H_{2}O}$ abundance ratio was also much lower than the value published in the \citet{Fink:2009} study. One possibility is that there was another parent molecule besides H$_{2}$O that contributed to the observed O(${^1}$D) 6300 {\AA} flux. This is explored by \citet{Decock:2013} who used intensity ratios of O(${^1}$S) and O(${^1}$D) emission lines from observations of 12 comets to deduce that at heliocentric distances greater than 2.5 au the contribution of CO$_{2}$ molecules to the [OI] flux increases (see their fig. 18). However, most of the comets in this work that had the [OI] emission present in their spectra were observed at r$_{h}$ < 2.5 au and were therefore in the heliocentric distance range where the production of oxygen atoms is mostly due to dissociation of H$_{2}$O molecules according to the \citet{Decock:2013} study. We therefore suggest that the low resolution of the spectrograph combined with the fact that the emission bands in the red arm spectra were not very strong above the dust continuum was the reason for not obtaining $\frac{NH_{2}}{H_{2}O}$ abundances that were consistent with the published average value. Another possibility is our  corrections to convert our through-the-slit flux to total luminosity are incorrect, although we cannot find an error that would account for such a large discrepancy.
We therefore discarded these measurements and do not report them here.

\subsection{Dust activity and colour}
To measure relative dust production rates we used the $Af\rho$ formalism of \cite{AHearn:1984a}. We integrated the total flux from the extracted flux calibrated spectra at $\lambda=4410-4490$ \AA \ (see below). The $Af\rho$ formalism assumes a canonical $\rho^{-1}$ surface brightness profile measured with a circular aperture of area $\pi\rho^{2}$, but the apertures used to extract the cometary fluxes were rectangular. We therefore used the transformation derived by \citet{AHearn:1984b} to obtain an effective radius $\rho$ of the circular aperture that would give the same filling factor $f$ as our rectangular apertures. Our measured fluxes then give a value for $Af\rho$ for this effective radius. The resulting dust activity measurements are given in Table \ref{tab:dust_slopes}. As almost all comets exhibited significant dust comae, we assumed that the contribution from the central nucleus was minimal. We also corrected these measurements to a phase angle of $0^\circ$ using the composite phase function published by \cite{Schleicher:2010b}. The exception to this procedure was P/2016 BA14, where imaging at this time revealed an extremely weak dust coma \citep{Li:2017}. Hence for this comet we assumed the contribution from the nucleus was non-negligible, and we did not attempt to measure $Af\rho$ for this object.

Dust comae have previously been observed to be generally red with respect to Solar colours, but with significant variations from comet to comet {\it e.g.} \cite{Storrs:1992}. We  measured the dust colour by dividing the flux calibrated comet spectra by the solar analogue star HD~28099 \citep{Hardorp:1980} obtained on the same night. For all comets we first normalised the blue arm reflectance spectrum at $\lambda=5260$ \AA. We then measured the reflectance in spectral regions $4450\pm40$ \AA \ and $5260\pm30$ \AA \ which are relatively free from gas emission in active comets. These are approximately equivalent to the BC and GC continuum filter bandpasses of the Hale-Bopp narrow band cometary filter set \citep{Farnham:2000}. For the red arm, we normalised the spectrum at $\lambda=5850$ \AA\ and used continuum regions at $5860\pm100$ \AA\ and $8600\pm100$ \AA, as these appeared relatively free from terrestrial absorption and cometary gas emission in the highly active comet 333P. Reflectance slopes were calculated via performing an optimal straight-line fit of these continuum regions in the blue and red arms using the {\tt scipy.optimize.curve\_fit} routines in \textsc{Python}, and are given in Table \ref{tab:dust_slopes}.

%Afp values and dust reflectance slopes.
\begin{table*}
	\centering
	\caption{$Af\rho$ relative dust production rates with effective aperture radii $\rho$,  and linear dust continuum reflectance slopes. Positive slopes indicate reflectance increasing with wavelength.}
	\label{tab:dust_slopes}
	\begin{tabular}{lccccc} 
		\hline
		Comet & $Af \rho$ & $Af \rho (0^\circ) $ & Effective $\rho$ & S'($4450-5260$ \AA ) & S'($5850-8650$ \AA ) \\
          & \multicolumn{2}{c}{(cm)} & (km) &  \multicolumn{2}{c}{(\%/$10^3$\AA )}\\
		\hline
		9P/Tempel 1 & 59.3 & 86.0 & 3370 & $12.6\pm 1.8$ & $9.8\pm 0.2$\\
        67P/Churyumov-Gerasimenko & 37.5 & 45.7 & 2460 & $15.4\pm2.9$ & $11.7\pm 0.4$\\
        77P/Longmore & 107 & 160 & 4630 & $19.9\pm 1.9$ & $10.5\pm 0.2$\\
       	81P/Wild 2 & 257 & 621 &  2110 &  $21.8\pm1.8$ & $9.9\pm 0.1$\\
     	116P/Wild 4 & 105 & 223 &  5180 & $14.8\pm1.9$ & $10.1\pm 0.3$\\
     	333P/LINEAR & 23.8 & 69.9 & 2992 & $22.2\pm 1.1$ & $3.5\pm0.2$\\
     	C/2013 US10 (Catalina)  & 904  & 2010 & 7220 &   $9.5\pm 1.5$ & $5.7\pm0.2$\\
		C/2014 S2 (PANSTARRS) & 2120 & 4390 &  6060 & $8.6\pm 1.6$ &  $5.6\pm 0.1$\\
       	C/2014 W2 (PANSTARRS) & 475  & 969 & 8840 & $20.3\pm 0.8$ & $4.8\pm 0.1$\\
        C/2015 V2 (Johnson) & 1700 & 2490 & 7716 &  $8.2\pm 1.5$ & $-0.8\pm 0.1$\\
        P/2016 BA14 (PANSTARRS) & $-$  &  $-$ & $-$ &  $-6.2\pm 1.0$ & $10.3\pm 0.1$\\
		\hline
	\end{tabular}
\end{table*}

\section{Discussion}
\subsection{Detection of the OH (0-0) emission band}
Five out of the twelve comets observed over two nights had the OH emission band at 3085 {\AA} present in their spectra. This demonstrates that the ISIS spectrograph is a viable instrument for measuring the emission from this molecule, and therefore determining the water ice sublimation rate. The effects of atmospheric extinction at La Palma are significant for the OH (0-0) flux, particularly for comets observed at high airmass values, as the extinction coefficient $k$ = 1.73 mag airmass$^{-1}$ at 3085 {\AA} \citep{King:1985}. From our data analysis and consideration of the reported magnitudes of the comets at the time of observation, we find that the limit for detection of the OH 3085 {\AA} emission band for a total exposure time of 1 hour using ISIS is an apparent coma magnitude of V${\sim}$15.

\subsection{\texorpdfstring{Measuring the C$_{2}$ production rates}{C2}}
During the initial analysis of our data, the measured log($\frac{C_{2}}{CN}$) values seemed unusually high in comparison to the average values determined by \citetalias{AHearn:1995} and \citetalias{Cochran:2012}. We also noted that the extracted C$_{2}$ profiles for most of the comets appeared to get significantly steeper at nucleocentric distances less than ${\sim}$10$^4$ km, in contrast to the expected behaviour of daughter species. The effect is much more significant in spectra that are of lower signal-to-noise. It is unlikely that all of the comets we observed belong to a category of objects that have enhanced C$_{2}$ in their composition, therefore we re-examined our data to investigate this. For each night, we divided our flux-calibrated solar analogue spectrum by a standard solar reference spectrum\footnote{https://www.nrel.gov/grid/solar-resource/spectra-astm-e490.html}. We found that the ratio spectrum was flat over most of the wavelength range as would be expected, but there was structure present in the wavelength region at $\lambda>5100$ \AA . In the cometary exposures this had the effect of `increasing' the flux in the C$_{2}$ ($\Delta$v=0) emission band, which meant that the derived production rate was also enhanced.  This structure was due to incomplete accounting for the strong dichroic response at $5100<\lambda<5500$ {\AA} \citep{Skillen:2016}.

Using the ratio of the extracted HD28099 spectrum and the standard solar spectrum we performed a fit to the structure in the C$_{2}$ wavelength region, and applied this correction to our extracted one-dimensional cometary spectra. As well as correcting the flux in the C$_{2}$ ($\Delta$v=0) emission band, the dust continuum in the region redward of the band also became significantly more flat. Using the corrected spectra we re-measured the integrated flux in the dust-subtracted C$_{2}$ ($\Delta$v=0) emission feature, and used this to compute C$_{2}$ production rates. The C$_{2}$ production rates in \cref{fig:s2_profiles_ahearn,fig:s2_profiles_cochran} are those obtained from the spatial profile fitting and is given for illustration purposes only. The C$_{2}$ production rates determined from the one-dimensional spectra after accounting for the effects of the dichroic are given in \cref{tab:ahearn_blue_prod_rate_table,tab:cochran_prod_rate_table}, and these values were used in all further abundance ratio calculations.  

\subsection{Abundance ratios determined using the A'Hearn et al. (1995) scale lengths}
\citetalias{AHearn:1995} define depleted comets as having log($\frac{C_{2}}{CN}$) < -0.18, and therefore 81P/Wild 2 would be classified as depleted based on our measurements. The log($\frac{C_{2}}{CN}$) value for  comet C/2013 US10 is close to the boundary between typical and depleted. The C$_{2}$ abundance measured for  comets 77P/Longmore and 116P/Wild 4 are high, although they are still within the range of values for typical comets found by \citetalias{AHearn:1995}. We also note that the CN spatial profiles extracted for this comet was noisy and the emission did not extend far from the nucleus, therefore the CN production rate for this comet was measured using the integrated band flux rather than a fit to the profile. The average C$_{2}$ abundances calculated excluding 81P and C/2013 US10 are also consistent with the average values for typical comets.

The average abundances of C$_{3}$ and NH relative to both OH and CN are all consistent with the average values for typical comets measured by \citetalias{AHearn:1995} despite the small number of comets that had NH emission detected in their spectra. \citetalias{AHearn:1995} note that there is no distinction between typical and depleted comets based on their $\frac{NH}{OH}$ ratios. The log($\frac{CN}{OH}$) abundance ratio measured by us is also consistent with the \citetalias{AHearn:1995} value. 

\subsection{Abundance ratios determined using the Cochran et al. (2012) scale lengths}
\citetalias{Cochran:2012} strictly define depleted comets as having both log($\frac{C_{2}}{CN}$) $\leq$ 0.02 and log($\frac{C_{3}}{CN}$) $\leq$ -0.86, but also include a depleted class that only have log($\frac{C_{2}}{CN}$) $\leq$ 0.02. Comets 81P/Wild 2 and C/2013 US10 would therefore be classified as depleted based on the stricter definition of \citetalias{Cochran:2012}. As discussed in the previous section, the C$_{2}$ abundance relative to CN for 77P/Longmore and 116P/Wild 4 seem high. The average log($\frac{C_{2}}{CN}$) and log($\frac{C_{2}}{OH}$) values calculated excluding  comets 81P and C/2013 US10 are consistent with the values for typical comets determined by \citetalias{Cochran:2012}.

The average log($\frac{C_{3}}{CN}$) value measured by \citetalias{Cochran:2012} for typical comets is different to the value in the \citetalias{AHearn:1995} study. This is due to the fact that \citetalias{Cochran:2012} used a different fluorescence efficiency for C$_{3}$ which had the effect of significantly increasing the computed column densities. We therefore scaled the extracted profiles using this adjusted fluorescence efficiency value when carrying out the Haser fitting for C$_{3}$ using the scale lengths from \citetalias{Cochran:2012}. This results in an average log($\frac{C_{3}}{CN}$) value that is less negative compared to that determined using the \citetalias{AHearn:1995} scale lengths. Our measured average log($\frac{C_{3}}{CN}$) value is consistent with the \citetalias{Cochran:2012} value for typical comets. However, our measured C$_{3}$ abundance relative to OH seems more similar to the \citetalias{Cochran:2012} value for depleted comets. Excluding the depleted comets in our data set from the calculation gives a log($\frac{C_{3}}{OH}$) value that is consistent with the \citetalias{Cochran:2012} average measured for typical comets.  

\begin{figure}
	\includegraphics[width=\columnwidth]{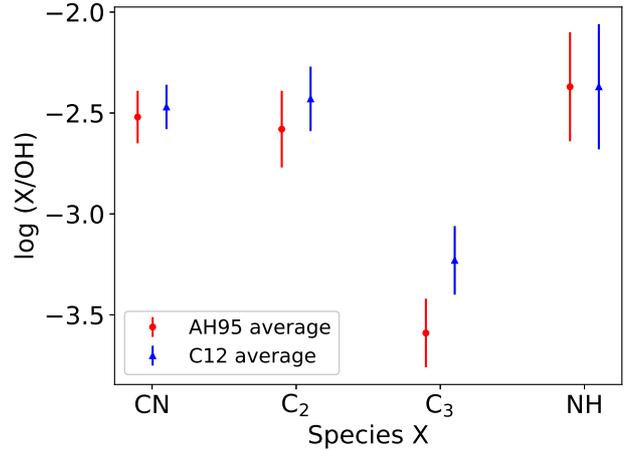}
    \caption{Comparison plot of average log ratios of each species X to OH.}
   \label{fig:oh_comparison}
\end{figure}

The average abundance of NH relative to both OH and CN is lower than expected from the published averages, but this could be due to the fact that NH profiles could only be extracted for three of the comets which limited the number of measurements used to calculate the overall average value. Our measured average log($\frac{NH_{2}}{OH}$) ratio is consistent within the error bars with the \citetalias{Cochran:2012} typical value, but the NH$_{2}$ abundance relative to CN appears to be higher than expected. \citetalias{Cochran:2012} measure NH$_{2}$ production rates using spatial profiles, but we were unable to extract profiles from our red arm observations. Instead, we determined the NH$_{2}$ production rates using the integrated band flux measured from the one-dimensional red spectra. We point the reader to Section 3.3, where we noted that the $\frac{NH_{2}}{H_{2}O}$ abundance ratio determined using the method of \citet{Fink:2009} was also inconsistent with the published value.    
In Fig.~\ref{fig:oh_comparison} we compare the average abundance ratios of each species relative to OH measured using the \citetalias{AHearn:1995} and the \citetalias{Cochran:2012} scale lengths, to demonstrate the effect of changing these parameters on the abundance ratios.

%Plot of average abundances relative to OH.

\subsection{Comparison with previous measurements}
In this section we compare our abundance ratio measurements with those of previous authors. We also compare our determined production rates for the various coma species with any previous measurements at similar $r_{h}$. We have not identified any previously published values for gas production rates in Comets 333P/LINEAR, C/2014 S2 and C/2014 W2.

\subsubsection{9P/Tempel 1}
9P/Tempel 1 is a Jupiter family comet, and is one of a small number of comets that have been observed in situ using spacecraft as it was the target of the Deep Impact mission \citep{AHearn:2005}. This object was included in the studies of \citetalias{AHearn:1995} and \citetalias{Cochran:2012}. Both sets of authors put this comet in the typical category, determining log($\frac{C_{2}}{CN}$) values of -0.090${\pm}$0.004 and 0.07${\pm}$0.01 respectively. \citet{Lara:2006} and \citet{Schleicher:2007} also suggest that 9P is a typical comet. However, \citet{Fink:2009} find 9P to be depleted in C$_{2}$, and define a `Tempel 1' class of comets that have low C$_{2}$ but normal NH$_{2}$ abundance. In this study we find a log($\frac{C_{2}}{CN}$) value of 0.06${\pm}$0.03 using the \citetalias{AHearn:1995} scale lengths and a value of 0.15${\pm}$0.03 using the scale lengths in \citetalias{Cochran:2012}. These measurements are consistent with the comet being of typical composition.

We observed 9P/Tempel 1 at a pre-perihelion heliocentric distance of 2.01 au, and \citetalias{Cochran:2012} measure production rates on several dates at a comparable pre-perihelion $r_{h}$ of 2.03 au. In 1983 February the CN production rate was 1.4$\times10^{24}$ \molsec, while in February 1994 the measured $Q_{CN}$ was 2.4$\times10^{24}$ \molsec. Using the \citetalias{Cochran:2012} scale lengths we measured a $Q_{CN}$ of (8.1$\pm$0.3)$\times10^{23}$ \molsec in March 2016, which is an apparent decrease compared to the earlier values. The $Q_{OH}$ value of 2.8$\times10^{26}$ \molsec measured by \citetalias{Cochran:2012} in 1994 February is comparable to our measured value of (3.3$\pm$0.6)$\times10^{26}$ \molsec. \citetalias{Cochran:2012} measured C$_{3}$ and C$_{2}$ production rates of 6.0$\times10^{23}$ \molsec and 3.5$\times10^{24}$ \molsec respectively in 1994 February, which are higher than our measured production rates for these species (see Table~\ref{tab:cochran_prod_rate_table}).

\citet{Schleicher:2007} used the scale lengths of \citetalias{AHearn:1995} to measure a C$_{2}$ production rate of (2.0$\pm$0.7)$\times10^{24}$ \molsec in 1983 February. Our measured $Q_{C_{2}}$ in Table~\ref{tab:ahearn_blue_prod_rate_table} is a factor 2.5 smaller than this. However, it is important to note that \citet{Schleicher:2007} assumed an outflow velocity $v=1$ km s$^{-1}$ independent of $r_h$,  but in this work we used the $r_h$ dependence given in equation \eqref{eq:2}.  In the Haser model the production rate of a particular species is proportional to the outflow velocity (see equation \eqref{eq:1}). Taking this into account, the \citet{Schleicher:2007} C$_{2}$ production rate becomes $Q_{C_2}= (1.2\pm0.4)\times10^{24}$ \molsec, which is within a factor 1.5 of our measured value.

\cite{Schleicher:2007} measured $Af\rho= 96$ cm in 1983 at a very similar heliocentric distance of $r_{h}=2.05$ au. Correcting to $0^\circ$ as for our data gives $Af\rho(0^\circ)= 197$ cm, over twice our extrapolated value. We note this previous study demonstrated both the existence of significant secular trends in gas and dust production for this comet, and also a coma dust distribution that deviated strongly from the canonical $1/\rho$ surface brightness distribution. As our calculation of an effective $Af\rho$ assumes a simple dust coma to transform from a spectroscopic slit to an effective circular aperture, it is dangerous to infer any secular variation in dust production rates from this comparison.

\subsubsection{67P/Churyumov-Gerasimenko}
67P/Churyumov-Gerasimenko is a Jupiter Family comet that was the target of the Rosetta mission. We were only able to measure the CN production rate for this comet at r$_{h}$=2.61 au post-perihelion. The CN spatial profile extracted from our observations for this comet was noisy and the emission did not extend far from the nucleus, therefore we determined the CN production rate using the integrated band flux. \citet{Opitom:2017} observed 67P with VLT/FORS2 on 2016 March 14 and 28 at post-perihelion heliocentric distances of 2.57 au and 2.67 au respectively, and measured $Q_{CN}$ values  of (1.05$\pm$0.11)$\times10^{24}$ \molsec and (1.17$\pm$0.16)$\times10^{24}$ \molsec using scale lengths from \citetalias{AHearn:1995}. Our measured $Q_{CN}$ given in Table~\ref{tab:ahearn_blue_prod_rate_table} is a factor of 2 smaller than these measurements. Scaling these values to include the $r_{h}^{-0.5}$ dependence for the outflow velocity in equation \eqref{eq:2} gives CN production rates of (5.52$\pm$0.58)$\times10^{23}$ \molsec and (6.16$\pm$0.84)$\times10^{23}$ \molsec, which are now only 1.1-1.2 times higher than our measured $Q_{CN}$ values.

\cite{Knight:2017} measured $Af\rho(0^\circ)=242$ cm 8 days before our observation, using Gemini-N+GMOS with a $g'$ filter. With a reflectance slope of $\sim 13$\%/1000\AA , this would correspond to $Af\rho(0^\circ)\sim232$cm at $r_h=2.54$ au at our measurement wavelength. \cite{Opitom:2017} report TRAPPIST $R_c$ band measurements of $Af\rho(0^\circ)=106$ cm at $r_h=2.52$ au and $Af\rho(0^\circ)=91.6$ cm at $r_h=2.67$ au. Interpolating and correcting for spectral slope would imply $Af\rho(0^\circ)\simeq 72$ cm with our measurement wavelength and epoch. Our measurement of $Af\rho(0^\circ)= 46$ cm is significantly lower than both of these measurements. It is more in line with the spectroscopic measurement of \cite{Ivanova:2017} one month later, whose $Af\rho = 46$ cm is similarly lower than imaging measurements made by the above authors near that time. We interpret this as additional evidence that values of $Af\rho$ measured via spectroscopy tend to fall below those measured in images, similar to the low value exhibited above for 9P, due to deviations from a canonical $\rho^{-1}$ dust surface brightness profile.

\subsubsection{77P/Longmore}
77P/Longmore is a Jupiter Family comet with very limited observations available in the literature. We were unable to find any published abundance ratios for this comet. \citetalias{Cochran:2012} observed this object in 1981 January and 1988 April at pre-perihelion heliocentric distances of 2.98 au and 2.65 au respectively, but did not detect any molecular emission in the spectra. The CN spatial profile extracted from our observations for this comet was noisy and the emission did not extend far from the nucleus, therefore we determined the CN production rate using the integrated band flux. The log($\frac{C_{2}}{CN}$) value is at the higher end of the range of values measured by \citetalias{AHearn:1995} for typical comets.

\subsubsection{81P/Wild 2}
81P/Wild 2 is a Jupiter family comet that was the target of the Stardust mission \citep{Brownlee:2004}. The measurements of the $\frac{C_{2}}{CN}$ abundance ratio for this comet suggest that it is in the depleted category (\citet{AHearn:1995,Farnham:2005,Fink:2009,Cochran:2012,Lin:2012}). In this study we find a log($\frac{C_{2}}{CN}$) value of -0.30${\pm}$0.03 using the \citetalias{AHearn:1995} scale lengths and a value of -0.20${\pm}$0.03 using the scale lengths in \citetalias{Cochran:2012}; these values would put this comet in the depleted category. The $\frac{C_{3}}{CN}$ ratio measured in this paper is also consistent with the average for the depleted class of comets.

Our observations were made when 81P/Wild 2 was at a pre-perihelion heliocentric distance of 1.99 au. \citetalias{Cochran:2012} observed this comet in 1984 April at an $r_h$ of 2.04 au pre-perihelion and measured a CN production rate of 1.0$\times10^{25}$ \molsec, which is ${\sim}$1.8 times larger than the $Q_{CN}$ of (5.5$\pm$0.1)$\times10^{24}$ \molsec that we measured in 2016 March. \citet{Farnham:2005} used the scale lengths from \citetalias{AHearn:1995} to measure the production rates of several molecules in 1997 January when 81P was at an $r_{h}$ of 1.91 au pre-perihelion. They measure $Q_{OH}$ = (4.2$\pm$0.1)$\times10^{27}$ \molsec, $Q_{CN}$ = (1.0$\pm$0.1)$\times10^{25}$ \molsec, $Q_{C_{3}}$ = (1.7$\pm$0.2)$\times10^{24}$ \molsec and $Q_{C_{2}}$ = (5.4$\pm$1.3)$\times10^{24}$ \molsec, which are higher than our measurements in Table~\ref{tab:ahearn_blue_prod_rate_table}. Scaling these values to include the $r_{h}^{-0.5}$ dependence for the outflow velocity in equation \eqref{eq:2} gives adjusted values of $Q_{OH}$ = (2.5$\pm$0.1)$\times10^{27}$ \molsec, $Q_{CN}$ = (6.0$\pm$0.6)$\times10^{24}$ \molsec and $Q_{C_{2}}$ = (3.3$\pm$0.8)$\times10^{24}$ \molsec. These are factors of 1.9, 1.3 and 1.4 higher than our OH, CN and C$_{2}$ production rate measurements respectively.

\cite{Farnham:2005} report $Af \rho\simeq500$ cm at $r_h=1.90$ au in 1997, equivalent to $Af \rho (0^\circ) \simeq530$ cm at zero phase angle. Our measurements imply a slightly higher dust production rate by $\sim 20$\% at $r_h=1.99$ au.

\subsubsection{116P/Wild 4}
116P/Wild 4 is a Jupiter family comet that has limited observations available in the literature. \citetalias{AHearn:1995} place this comet in the depleted category based on its $\frac{C_{2}}{CN}$ value, but it is not included in their restricted data set as there are only two observations of this object in their study. This comet was not observed in the survey carried out by \citetalias{Cochran:2012}. \citet{Fink:1996} note that no emission from this comet was seen at a heliocentric distance of 2.19 au and only weak activity was observed at 2.01 au. We measure a log($\frac{C_{2}}{CN}$) of 0.23${\pm}$0.05 for this comet using the \citetalias{AHearn:1995} scale lengths, which is at the higher end of the range of measured values for typical comets. 

\citetalias{AHearn:1995} report $Af \rho\simeq210$ cm at $r_h=1.90$ au in 1990, equivalent to $Af \rho (0^\circ)  \simeq400$ cm, approximately twice our value.

\subsubsection{C/2013 US10 (Catalina)}
C/2013 US10 is a long-period comet that reached perihelion on 2015 November 15. \citet{Combi:2018} have published water production rates that were determined using SOHO/SWAN observations of hydrogen Lyman-$\alpha$. We observed this comet at a post-perihelion distance of 2.21 au, and the power law fit in the \citet{Combi:2018} study yields a water production rate of 3.6$\times10^{28}$ \molsec at this distance. Using the OH (0-0) emission observed in the spectrum of this comet and assuming a 90\% branching ratio for H$_{2}$O photodissociation \citep{Crovisier:1989}, we derive a water production rate of (8.8$\pm$1.7)$\times10^{27}$ \molsec. This water production rate is  a factor of ${\sim}$4 smaller than expected from the reported post-perihelion activity curve. We note that the observations in the \citet{Combi:2018} paper do not extend to the post-perihelion heliocentric distance where we observed C/2013 US10. Therefore a possible reason for the discrepancy is that the derived power law for $Q(OH)$ in this comet was not valid at $r_{h}$ = 2.21 au.

\cite{Protopapa:2018} report near-IR photometry at $3.4\mu$m giving a large value of $Af\rho=2500$ cm at $r_h=1.8$ au. Aside from the large difference in $r_h$, it is likely that the dust phase scattering function is different at these wavelengths, making comparison problematic. However our measurements also imply a large amount of dust associated with this comet.

\subsubsection{C/2014 S2 (PANSTARRS)}
This comet had the largest dust production rate measured by us, but we are not aware of any other $Af\rho$ measurements being reported for this comet. Although $Af\rho$  is large, we note that it is still an order of magnitude lower than exhibited by C/1995 O1 (Hale-Bopp) at a similar heliocentric distance \citep{Weiler:2003}.

\subsection{Dust reflectance properties}
The dust reflectance slopes in Table \ref{tab:dust_slopes} are mostly similar to those previously measured. For example, \cite{Storrs:1992} found an average slope of $<S'>=22\pm7$\%/1000 \AA \ over $\lambda=4400-5650$ \AA\ from measurements of 19 comets, and \cite{Jewitt:1986} found $<S'>=13\pm5$\%/1000 \AA \ over $\lambda=3500-6500$ \AA\ from measurements of 9 comets. Our blue arm spectra give $<S'>=13\pm8$\%/1000 \AA, in agreement with these previous observations. It is also clear from our data that almost all comets become less red as wavelength increases. This general decrease in reflectance slope with wavelength was first investigated by \cite{Jewitt:1986} and is observed for most cometary dust. 

A clear outlier in our data is the blue arm reflectance slope for P/2016 BA14. Inspection of this spectrum showed a decrease in reflectance beyond $\lambda\sim5000$ \AA \ in the blue arm, but the red arm exhibited the expected increase in reflectance at wavelengths $\lambda > 5400$ \AA . Given the previous issues with our data in this spectral region described in section 4.2, we do not assign any significance to this measurement. However we did not see a similar issue with any of the other comets, and suggest that this would be an interesting object to re-observe in a future apparition.

Discounting this object, the most notable object is C/2015 V2. A neutral/blue spectral slope is common at near-infrared wavelengths \citep{Jewitt:1986}, but it is highly unusual at red optical wavelengths. This was the most distant comet observed by us, but as no gas emissions were detected by us or have been reported elsewhere, it is unknown if this comet was unusual in other aspects. Finally, we find no correlation between dust colour and orbital properties for our comets, agreeing with previous studies {\it i.e.} \cite{Solontoi:2012}.

\section{Conclusions}
One of the aims of this work was to investigate the capability of the ISIS spectrograph to detect OH (0-0) emission, as the majority of other ground-based spectrographs have a wavelength cut-off that is redward of this band. We observed eleven comets in 2016 March and detected the OH (0-0) emission band in the spectra of five of these objects. From the analysis of the data we conclude that observations of comets brighter than V${\sim}$15.0 made using ISIS allows accurate OH production rates with a total exposure time up to one hour. Of course, fainter comets may be observed with longer integration times or at low airmass, and the total OH in the slit will also depend on the Earth-comet geometry.

We also observed emission bands due to several other species in the extracted spectra, with the blue arm containing emission bands of NH, CN, C$_{3}$ and C$_{2}$. The red arm contained emission bands of NH$_{2}$, [OI] and CN, however the strengths of these features above the dust continuum were weak in comparison to those in the blue arm. We suggest that future observations of gas using the red arm of ISIS should only be carried out on relatively bright comets, or at higher spectral resolution.

Using the blue arm observations, we were able to determine production rates and abundance ratios for various species using Haser modelling. We chose to use the scale lengths for the parent and daughter species used by previous authors rather than determining them from our own dataset, as this allows comparison of our measured abundances with other values in the literature. We found that our measured average abundance ratios were generally consistent with those published by \citetalias{AHearn:1995} and \citetalias{Cochran:2012} despite our limited number of measurements compared to those larger studies. Finally, we found a trend for our $Af\rho$ measurements to be lower than those obtained through imaging studies. This may be due to a deviation from a canonical $1/\rho$ dust distribution in the observed comets.

\section*{Acknowledgements}
We are grateful to the referee Anita Cochran whose comments improved the quality and clarity of this manuscript.
The William Herschel Telescope is operated on the island of La Palma by the Isaac Newton Group of Telescopes in the Spanish Observatorio del Roque de los Muchachos of the Instituto de Astrof\'{i}sica de Canarias.
We gratefully thank the staff at the Isaac Newton Group for their support in the observations reported here.
MGH acknowledges receipt of a Northern Ireland DEL studentship award. AF acknowledges support for this work from UK STFC grants ST/L000709/1 and ST/P0003094/1. CS acknowledges support from the STFC in the form of the Ernest Rutherford Fellowship.

%The Acknowledgements section is not numbered. Here you can thank helpful
%colleagues, acknowledge funding agencies, telescopes and facilities used etc.
%Try to keep it short.

%%%%%%%%%%%%%%%%%%%%%%%%%%%%%%%%%%%%%%%%%%%%%%%%%%

%%%%%%%%%%%%%%%%%%%% REFERENCES %%%%%%%%%%%%%%%%%%

% The best way to enter references is to use BibTeX:

\bibliographystyle{mnras}
\bibliography{paper_references} 

%%%%%%%%%%%%%%%%%%%%%%%%%%%%%%%%%%%%%%%%%%%%%%%%%%

%%%%%%%%%%%%%%%%% APPENDICES %%%%%%%%%%%%%%%%%%%%%

%\appendix

%\section{Some extra material}

%If you want to present additional material which would interrupt the flow of the main paper,
%it can be placed in an Appendix which appears after the list of references.

%%%%%%%%%%%%%%%%%%%%%%%%%%%%%%%%%%%%%%%%%%%%%%%%%%

% Don't change these lines
\bsp	% typesetting comment
\label{lastpage}
\end{document}